\documentstyle[aps,multicol]{revtex}

\newcommand{\ket}[1]{\vert #1 \rangle}

\begin{document}
\draft
\title{Phase randomization improves the security of quantum key distribution}
\author{Hoi-Kwong Lo,$^1$ and John Preskill$^2$}
\address{
$^1$ Department of Electrical and Computer Engineering and Department of Physics, \\University of Toronto, Toronto, Canada M5G 3G4\\
$^2$ Institute for Quantum Information, California Institute of Technology, 
Pasadena, CA 91125, USA
}
\maketitle
\begin{abstract}

Ideal quantum key distribution (QKD) protocols call for a source that emits single photon signals, but the sources used in typical practical realizations emit weak coherent states instead. A weak coherent state may contain more than one photon, which poses a potential security risk. QKD with weak coherent state signals has nevertheless been proven to be secure, but only under the assumption that the {\em phase} of each signal is {\em random} (and completely unknown to the adversary). Since this assumption need not be fully justified in practice, it is important to know whether phase randomization is really a requirement for security rather than a convenient technical assumption that makes the security proof easier. Here, we exhibit an explicit attack in which the eavesdropper exploits knowledge of the phase of the signals, and show that this attack allows the eavesdropper to learn every key bit in a parameter regime where a protocol using phase-randomized signals is provably secure. Thus we demonstrate that phase randomization really does enhance the security of QKD using weak coherent states. This result highlights the importance of a careful characterization of the source for proofs of the security of quantum key distribution.


\end{abstract}
 
\begin{multicols}{2}

In quantum key distribution \cite{BB84}, two parties (Alice and Bob) use quantum signals to establish a shared key that can be used to encrypt and decrypt classical messages. An eavesdropper (Eve) who collects information about the key by interacting with the signals produces a detectable disturbance; therefore Alice and Bob can detect the eavesdropper's activity, and they can reject the key if they fear that the eavesdropper knows too much about it. But if the detected disturbance is weak enough, then Alice and Bob can use classical error correction and privacy amplification protocols to extract a shared key that is very nearly uniformly distributed and almost certainly private \cite{mayers,lo-chau,biham,ShorPreskill}. The security of the QKD protocol is said to be {\em unconditional}, because the security can be proven for any attack consistent with the laws of quantum physics, and without any assumptions about computational hardness. 

Experiments have recently demonstrated the feasibility of QKD over 150 km telecom fibers \cite{NEC,GYS}, and at least two firms are now marketing commercial QKD systems \cite{commercial}. But how secure are these systems, really? To assess the security of practical implementations of QKD, it is important to consider how well the actual systems match the performance assumed in the security proofs. In particular, the signals used in typical practical realizations of QKD are dim laser pulses, which occasionally contain more than one photon. Multi-photon signals together with loss in the optical fiber can threaten security, but proofs of security for QKD using weak coherent states have been found\cite{ilm,GLLP}.
(We note that for QKD protocols that use decoy states \cite{Hwang,Decoy,Wang,practical}, security can be proven even for rather strong coherent-state signals. In this paper, however, we will focus on QKD with weak coherent states.)

A key assumption in the security proofs in \cite{ilm,GLLP} (and also in \cite{Decoy}) is that the phase of the quantum signal is uniformly random. A coherent state of one mode of the electromagnetic field can be expressed as
\begin{equation}
|\alpha\rangle = e^{-|\alpha|^2/2}\sum_{n=0}^\infty \frac{\alpha^n}{\sqrt{n!}} |n\rangle~,
\end{equation}
where $|n\rangle$ denotes the state with photon number $n$. We may write $\alpha=\sqrt{\mu}e^{i\theta}$, where $\mu=|\alpha|^2$ denotes the mean photon number and $e^{i\theta}$ is the {\em phase} of the coherent state. To an eavesdropper with no {\it a priori} knowledge of the phase, a signal whose phase is selected uniformly at random is indistinguishable from the state
\begin{equation}
\rho_\mu = \int_0^{2\pi} \frac{d\theta}{2\pi} |\sqrt{\mu}e^{i\theta}\rangle\langle 
\sqrt{\mu}e^{i\theta}|=e^{-\mu}\sum_n  \frac{\mu^n}{n!} |n\rangle\langle n|~,
\end{equation}
a Poisson distributed mixture of photon number eigenstates. Therefore, for a security analysis, we may suppose that a source emitting weak coherent state signals is actually emitting signals in the state $\rho_\mu$.

With probability $p_0=e^{-\mu}$, which is close to one for small $\mu$, the source emits no photon; exactly one photon is emitted with probability $p_1=\mu e^{-\mu}$. The probability that two or more photons are emitted is
\begin{equation}
p_M= 1- e^{-\mu}\left(1+\mu\right))\le \frac{1}{2} \mu^2~.
\end{equation}
Multiphotons can pose a security risk, but if each signal has a random phase, $p_M$ is sufficiently small, and the loss in the channel is not too high, then it is possible to prove security against arbitrary eavesdropping attacks \cite{ilm,GLLP}.

However, no known security proof applies if the eavesdropper has some {\em a priori} knowledge about the phase of the signal states. Conceivably, such phase information might be accessible in realistic implementations of QKD. For example, in a ``plug-and-play'' scheme, a strong signal is sent from Bob to Alice, who attenuates and modulates the signal before returning it to Bob; in unidirectional schemes as well, strong ancillary pulses are sometimes used to monitor the channel. The phase of a strong pulse is accurately measurable in principle, and could be correlated with the phase of the key-generating pulse. Even if strong pulses are not used, the phase coherence of a realistic source might be maintained during the emission of many weak signals, allowing the phase to be determined accurately. Therefore, it is important to investigate the security of QKD under the assumption that the eavesdropper knows something about the phase of the signals. Can she exploit this knowledge to improve the effectiveness of her attack?

Our conclusion is that she can. We will compare the security of two implementations of the BB84 QKD protocol \cite{BB84}, using two different sources. Source R (for random) emits phase randomized weak coherent states with mean photon number $\mu$. Source P (for phase) emits weak coherent states with the same mean photon number $\mu$, but such that each signal has a definite phase (e.g., $e^{i\theta}=1$) that is known by the eavesdropper. We will show that there is a range for the bit error rate $\delta$ observed in the protocol ($.146 < \delta < .189$) such that, for sufficiently small $\mu$, BB84 using source R is secure, but BB84 using source P is not. Specifically, using source R, key bits can be generated at a positive asymptotic rate such that, with high probability, the key is exponentially close to uniformly random and the eavesdropper's knowledge of the key is exponentially small. But if source P is used, for the same bit error rate and signal strength, the eavesdropper has perfect knowledge of every key bit.

Briefly, our argument goes as follows: We will describe an intercept/resend attack on BB84 using a measurement that we call {\em unambiguous key discrimination}. By exploiting her knowledge of the phase of the signals emitted by source P, Eve performs a POVM with three outcomes: 0, 1, and DK (don't know). The DK outcome is inconclusive, but if either of the other outcomes occurs, then Eve knows with certainty the key bit (0 or 1) encoded in the BB84 signal emitted by the source, though she does not gain any information about which of the two possible BB84 states compatible with that key bit was emitted. Eve blocks the signals when her outcome is DK, but if the outcome is conclusive she sends on to Bob a uniform coherent superposition of the two compatible BB84 states. This procedure generates a bit error rate $\delta = \frac{1}{2} - \frac{1}{2\sqrt{2}}\approx .146$. Evidently, Eve has the same key information as Alice and Bob, so that, if she also knows their protocol for error correction and privacy amplification, she will have perfect knowledge of every bit of the final key. On the other hand, if source R is used instead, by combining techniques in \cite{GLLP} with the two-way privacy amplification protocol in \cite{twoway}, we will show that for the same bit error rate, secure key can be extracted at a nonzero rate for $\mu < .0240$. Our essential observation is that while a fraction $O(\mu^2)$ of all signals are insecure when source R is used, this fraction increases to $O(\mu)$ when source P is used instead. Thus, QKD using source P is intrinsically more vulnerable to eavesdropping.

In the rest of this paper, we explain our argument in more detail. In the ideal (polarization-based) BB84 protocol, each signal is a single photon, and the key information is carried by the photon's polarization, a qubit. We consider two conjugate orthonormal bases for this qubit: the $z$ basis $\{|0\rangle,|1\rangle\}$ and the $x$ basis $\{|+\rangle,|-\rangle\}$, where $|\pm\rangle=\left(|0\rangle\pm |1\rangle\right)/\sqrt{2}$. The source emits one of these four states, chosen equiprobably. For signals sent in the $z$ basis, 0 and 1 are the key bits. For signals sent in the $x$ basis, $+$ indicates the key bit 0 and $-$ indicates 1. 

If the source actually emits weak coherent states, then the BB84 signals become
\begin{eqnarray} \label{bb84states}
\ket{\tilde 0} &=& e^{-\mu/2} \left(\ket{\rm vac} + \alpha \ket{0} +\dots \right)~, \nonumber \\
\ket{\tilde 1} &=& e^{-\mu/2} \left(\ket{\rm vac} + \alpha \ket{1} +\dots \right)~, \nonumber \\
\ket{\tilde +} &=& e^{-\mu/2} \left(\ket{\rm vac} + \alpha \ket{+} +\dots \right)~, \nonumber \\
\ket{\tilde -} &=& e^{-\mu/2} \left(\ket{\rm vac} + \alpha \ket{-} +\dots \right) ~,
\end{eqnarray}
where $|{\rm vac}\rangle$ denotes the vacuum (no-photon) state, and the ellipsis indicates the multiphoton contribution. Let us suppose that the phase of each signal is $e^{i\theta}=1$, so that $\alpha=\sqrt{\mu}$ is real and positive, We assume that $\alpha$ is small, so that multiphotons are unlikely. Ignoring the small multiphoton component, the signals reside in a qutrit Hilbert space with basis $\{|{\rm vac}\rangle,|0\rangle,|1\rangle\}$; expanded in this basis, they may be re-expressed as
\begin{eqnarray} \label{bb84vectors}
|\tilde 0\rangle& = & e^{-\mu/2}\left( 1, ~\alpha , ~~0 \right) ~,\nonumber \\
|\tilde +\rangle & = & e^{-\mu/2}\left( 1, { \alpha \over \sqrt{2} } , { \alpha \over \sqrt{2} } \right) ~,\nonumber \\
|\tilde 1\rangle& = & e^{-\mu/2} \left( 1,~~ 0, ~~\alpha \right) ~,\nonumber \\
|\tilde -\rangle & = & e^{-\mu/2} \left( 1, { \alpha \over \sqrt{2} } , -{ \alpha \over \sqrt{2} } \right)~.
\end{eqnarray}
We note that the four BB84 states span a plane in the Bloch sphere that can be chosen arbitrarily; thus we could replace the $x$ basis states by the rotated states $|\pm\rangle_\varphi=\left(e^{i\varphi}|0\rangle \pm e^{-i\varphi}|1\rangle\right)/\sqrt{2}$. In the ideal protocol the phase $e^{i\varphi}$ has no significance, but for weak coherent signals that are not phase randomized, the effectiveness of Eve's attack can depend on $e^{i\varphi}$. In eq.~(\ref{bb84vectors}) we have chosen $e^{i\varphi}=1$, because this choice minimizes Eve's ability to discern the value of the key bit. 

The multiphoton component of the state can help Eve, but to keep our analysis simple, we will consider an attack on the source P that makes no use of the multiphotons. Eve performs an orthogonal measurement that distinguishes photon number less than two from photon number greater than or equal to two, and she discards the state if the latter outcome is found. (We will be interested in values of $\mu$ that are sufficiently small that Eve would not benefit very much from taking advantage of the multiphotons.) Thus the states she retains are qutrits. She then performs a three-outcome POVM (unambiguous key distribution) to identify the key bit. The two conclusive outcomes of the POVM are projections onto the states:
\begin{eqnarray} \label{conclusive}
|0^\perp\rangle &=& N_{\mu,0} \left( -\alpha -{\alpha \over \sqrt{2}} ,~
1+ { 1 \over \sqrt{2}} ,~ {1  \over \sqrt{2}} \right) ~,\nonumber \\
|1^\perp\rangle &=& N_{\mu,1} \left( -{\alpha \over \sqrt{2}} ,~
1 +{ 1 \over \sqrt{2}} , ~{1  \over \sqrt{2}} \right) ~,
\end{eqnarray}
where $N_{\mu,0}, N_{\mu,1}$ are normalization factors such that 
\begin{eqnarray}
N_{\mu,0}^{-2}= (2+\sqrt{2})\left[1+\left(\frac{1}{2} +\frac{1}{2\sqrt{2}}\right)\mu\right]~,\nonumber\\
N_{\mu,1}^{-2}= (2+\sqrt{2})\left[1+\left(\frac{1}{2} -\frac{1}{2\sqrt{2}}\right)\mu\right]~.\nonumber\\
\end{eqnarray}
The vector $|0^\perp\rangle$ is orthogonal to both of the two states $|\tilde 0\rangle$ and $|\tilde +\rangle$ that indicate the key bit 0. Hence, if this outcome is found, Eve knows for sure that the key bit could not be 0 and so must be 1. Similarly, the vector $|1^\perp\rangle$ is orthogonal to both of the states $|\tilde 1\rangle$ and $|\tilde -\rangle$ that indicate the key bit 1. 

The vectors $|0^\perp\rangle$ and $|1^\perp\rangle$ are nearly parallel for small $\alpha$. To ensure that all three POVM elements are positive, we may choose 
\begin{eqnarray}
&&E_0= \frac{1}{2}|1^\perp\rangle\langle 1^\perp|~,\quad
E_1=\frac{1}{2}|0^\perp\rangle\langle 0^\perp|~,\nonumber\\
&&E_{\rm DK} = I - E_0 - E_1~.
\end{eqnarray}
(For small positive $\mu$, the strength of the conclusive POVM elements can be pushed up slightly, but this is a small effect that we ignore.) 
Thus we find the probability $p_D$ of a conclusive outcome (taking into account that Eve might detect multiphotons and reject the state)
\begin{eqnarray}
&&\langle \tilde 0|E_0|\tilde 0\rangle = \langle \tilde +|E_0|\tilde +\rangle \nonumber\\
&&= \left(\frac{1}{2} -\frac {1}{2\sqrt{2}}\right) \mu e^{-\mu}\left[1+\left(\frac{1}{2} +\frac{1}{2\sqrt{2}}\right)\mu\right]^{-1}
\end{eqnarray}
if the key bit is 0, and 
\begin{eqnarray}
&&\langle \tilde 1|E_1|\tilde 1\rangle = \langle \tilde -|E_1|\tilde -\rangle \nonumber\\
&&= \left(\frac{1}{2} -\frac {1}{2\sqrt{2}}\right) \mu e^{-\mu}\left[1+\left(\frac{1}{2} -\frac{1}{2\sqrt{2}}\right)\mu\right]^{-1}
\end{eqnarray}
if the key bit is 1. We note that the conclusive outcome is slightly more likely when the key bit is 1. This asymmetry can be traced to the property that the overlap $|\langle\tilde +|\tilde 0\rangle|$ of the two signals that indicate the key bit 0 is slightly larger than the overlap $|\langle\tilde -|\tilde 1\rangle|$ of the two signals that indicate the key bit 1. (For other choices of the phase $e^{i\varphi}$ that determines the plane in the Bloch sphere occupied by the BB84 signals, the asymmetry is substantially larger.) In any case, for either value of the key bit, the probability of a conclusive outcome obeys
\begin{equation}
p_D \ge (.146)\mu e^{-\mu}\left[1+(.854)\mu\right]^{-1}~.
\end{equation}

Now $p_D$ is the probability that Eve resends the signal to Bob, and therefore it is the probability that Bob detects a signal, if his detector is perfectly efficient. If there were no interference by the eavesdropper (and no loss in the quantum channel connecting Alice and Bob), all non-vacuum signals would be detected, and then $p_D=1-e^{-\mu}=\mu + O(\mu^2)$. If Eve uses the unambiguous key discrimination POVM, then of course $p_D$ vanishes in the limit $\mu\to 0$, but what is noteworthy (and crucial for our argument) is that $p_D$ vanishes {\em linearly} with $\mu$. Thus for $\mu$ small, a fraction $\eta\approx .146$ of order one of all the non-vacuum signals sent by Alice are received by Bob.

Having characterized Eve's attack when source P is used, we now consider the security analysis for source R, following \cite{GLLP}. We suppose that the source emits a multiphoton signal with probability $p_M$, that Bob detects a fraction $p_D$ of all the signals sent by Alice, and that Eve's attack is unrestricted. Of the signals that are received, the fraction that were emitted as multiphotons is no more than $\Delta=p_M/p_D$; the rest are single photon signals. 

We can prove security as in \cite{ShorPreskill}, by relating the BB84 protocol to a protocol is which the key is generated by measuring noisy entangled pairs shared by Alice and Bob. Private key can be extracted at a positive asymptotic rate if it is possible to distill high fidelity entanglement from the noisy entanglement. Entanglement distillation will succeed if the noisy entangled pairs have a bit error rate and a phase error rate that are both sufficiently small. The bit error rate $\delta$ is inferred directly from the verification test in the BB84 protocol; the phase error rate $\delta_p$ is also inferred, but by a less direct argument.

If the source and detector used in the protocol were perfect, then a symmetry argument would suffice to show $\delta=\delta_p$. This symmetry is broken if the equipment is imperfect, but it is still possible to bound the difference between the two error rates using an appropriate characterization of the imperfections. For the case where Bob has a perfect detector, but Alice's source sometimes emit multiphotons, it can be shown that 
\begin{equation}
|\delta_p-\delta| < \Delta/2~,
\end{equation}
where $\Delta$ is the fraction of all the detected signals that were emitted as multiphotons. (The upper bound $\Delta$ found in \cite{GLLP} has been improved to $\Delta/2$ in \cite{boileau}.)

Now, the security proof in \cite{ShorPreskill}, which relates one-way privacy amplification to one-way entanglement distillation using quantum error-correcting codes, does not apply for a bit error rate above $\delta=.110$. But security of BB84 was established in \cite{twoway} for a bit error rate as high as $.189$, by relating two-way privacy amplification to entanglement distillation with two-way communication between Alice and Bob. The original argument in \cite{twoway} assumed a perfect source and detector. But the two-way entanglement distillation succeeds if both $\delta$ and $\delta_p$ are below $.189$; therefore the argument can be applied to a protocol with imperfect equipment if there is a strong enough bound on $|\delta-\delta_p|$.

If the bit error rate $\delta$ is .146, then the two-way BB84 protocol is secure for $|\delta-\delta_p| < .189 - .146 = .043$. And for a source that emits phase-randomized coherent states, it suffices that $\Delta < .086$, where 
\begin{eqnarray}
\Delta= \frac{p_M}{p_D} &\le& \left(\frac{\frac{1}{2}\mu^2}{(.146)\mu e^{-\mu}\left[1+(.854)\mu\right]^{-1}}\right)\nonumber\\
&=&(3.42)\mu e^\mu\left[1+(.854)\mu\right]~.
\end{eqnarray}
Thus $\Delta < .086$, and the protocol is provably secure, for $\mu < .0240$. The security proof still applies if Bob's detector, rather than being perfectly efficient, has an efficiency that is independent of the basis in which the detector measures, where whether the detector fires is decided randomly, uninfluenced by the eavesdropper.

We have shown, therefore, that the BB84 QKD protocol is less secure using the phase-coherent source P than using the phase-randomized source R. Eve can exploit her knowledge of the phase of the signals emitted by the source P to implement a POVM that, when its outcome is conclusive, unambiguously identifies the key bit. But for the same bit error rate $\delta\approx.146$, signal strength $\mu$ ($< .0240$), and signal detection rate $p_D\approx .146 \mu$, if the signals have random phases then Alice and Bob can generate a final key about which Eve has negligible knowledge. 

This observation raises many questions. Can realistic sources be engineered so that each signal has a phase that the eavesdropper is unable to guess? Can ``plug-and-play'' systems, and other systems that use strong ancillary pulses, be protected from phase-coherent attacks? Can strong security results be proven if the phase is governed, not by a uniform probability distribution, but by a distribution that is sufficiently broad? Finally, and most urgently, if Eve knows the phase of every signal emitted by the source in a BB84 protocol, is there {\em any} positive bit error rate $\delta$ such that provably secure key can be generated at a positive asymptotic rate?

We thank Daniel Gottesman, Jeff Kimble, Norbert L\"{u}tkenhaus, Bing Qi, John Sipe, and Kiyoshi Tamaki for enlightening discussions. This work has been supported in part by: the Department of Energy under Grant No. DE-FG03-92-ER40701,  the National Science Foundation under Grant No. EIA-0086038,  the Caltech MURI Center for Quantum Networks under ARO Grant No. DAAD19-00-1-0374, NSERC, Canada Research Chairs Program, Canadian Foundation for Innovation, Ontario Innovation Trust,
Premier's Research Excellence Award, and Canadian Institute for Photonics Innovations. H.-K. Lo gratefully acknowledges the hospitality of the Institute for Quantum Information at Caltech, where this work was done.


\end{multicols}
\end{document}